\documentclass[11pt,twoside]{article}
\usepackage{asp2010}
\usepackage{graphicx}

\resetcounters

\begin{document}

\title{MHD simulations of a supernova-driven ISM and the warm ionized
  medium using a positivity preserving ideal MHD scheme}
\author{Mordecai-Mark Mac Low$^{1,4,9}$, Alex S. Hill$^{2,3}$, M. Ryan
  Joung$^4$, Knut~Waagan$^5$, Christian Klingenberg$^6$, Kenneth
  Wood$^7$, Robert~A.~Benjamin$^8$, Christoph Federrath$^{9,10,11}$, and
  L. Matthew Haffner$^2$
\affil{$^1$Dept.\ of Astrophysics, American Museum of Natural History,
79th St.\ at Central Park W., NY, NY, 10024, USA}
\affil{$^2$Dept.\ of Astronomy, U. Wisconsin-Madison, Madison, WI, USA}
\affil{$^3$CSIRO Astronomy \& Space Science, Marsfield, NSW, Australia}
\affil{$^4$Dept.\ of Astronomy, Columbia U., NY, NY, USA}
\affil{$^5$CSCAMM, U. Maryland at College Park, College Park, MD, USA}
\affil{$^6$Dept.\ of Mathematics, U. W\"urzburg, W\"urzburg, Germany}
\affil{$^7$School of Physics \& Astronomy, St. Andrews U.,
  St. Andrews, Scotland, UK}
\affil{$^8$Dept.\ of Physics, U. Wisconsin-Whitewater, Whitewater, WI,
  USA}
\affil{$^9$ZAH, Inst.\ f\"ur Theoretische Astrophysik, U. Heidelberg,
  Heidelberg, Germany}
\affil{$^{10}$CRAL, ENS Lyon, Lyon, France}
\affil{$^{11}$Monash Centre for Astrophysics, School of Mathematical Sciences, Monash U., Victoria, Australia}
}
\begin{abstract}
We present new 3D magnetohydrodynamic (MHD) simulations of a
supernova-driven, stratified interstellar medium.  These simulations
were run using the \citet{waagan11} 
positivity preserving scheme
for ideal MHD implemented in the Flash code. The scheme is stable even
for the Mach numbers approaching 100 found in this problem. We have
previously shown that the density distribution arising from
hydrodynamical versions of these simulations creates low-density
pathways through which Lyman continuum photons can travel to heights
$|z| > 1$ kpc. This naturally produces the warm ionized medium through
photoionization due primarily to O stars near the plane. However, our
earlier models reproduce the peak but not the width of the observed
emission measure distribution. Here, we examine whether inclusion of
magnetic fields and a greater vertical extent to the simulation domain
produce a gas distribution that better matches the observations. We
further study the change of magnetic energy over time in our models,
showing that it appears to reach a steady state after a few hundred
megayears, presumably supported by a turbulent dynamo driven by the
supernova explosions.
\end{abstract}

\section{Observational Motivation}

Supernova explosions provide one of the major sources of energy for
observed turbulent motions in the interstellar gas.  It can be argued that in
star forming regions of galaxies their influence predominates
\citep{maclow04,tamburro09}.  To constrain their influence on the
interstellar gas, we have compared observations of the upper
atmosphere and gaseous halo of the Galaxy to numerical simulations of
supernova-driven turbulence.  

The base model that we used, described by \citet{joung06} and
\citet{joung09}, and inspired by the work of \citet{avillez00}, used
version 2.4 of the Flash code \citep{fryxell00} to model a square
section through a stratified galactic disk centered on the midplane,
with dimensions $0.5 \times 0.5 \times 10$~kpc and periodic boundary
conditions in the horizontal direction. The vertical boundary
conditions were taken to be constant pressure to allow outflow.  The
adaptive mesh refinement capability of Flash was used to fix the
maximum resolution within 200~pc of the midplane to 1.95~pc, with
reduced maximum resolution at higher altitudes for computation
efficiency. Supernova explosions were placed at both random and
correlated locations to reproduce their distribution in the Galaxy.
Radiative cooling was implemented with an equilibrium ionization,
solely temperature dependent cooling function, and a diffuse heating
term characteristic of photoelectric heating and stratified with
altitude was also included.

\citet{henley10} measured X-ray emission from the hot, gaseous halo of
the Galaxy by using shadowing against relatively nearby clouds to
remove contributions from local sources including solar wind charge
exchange and emission from the Local Bubble. They then compared these
observations to simulations of emission from our models.  They found
good agreement between the simulated and observed emission measure of
hot gas looking out of the Galaxy.  However, the effective temperature
of the emission in the model systematically climbed with time.

\begin{figure}[t]
\centering
\includegraphics[scale=0.6
]{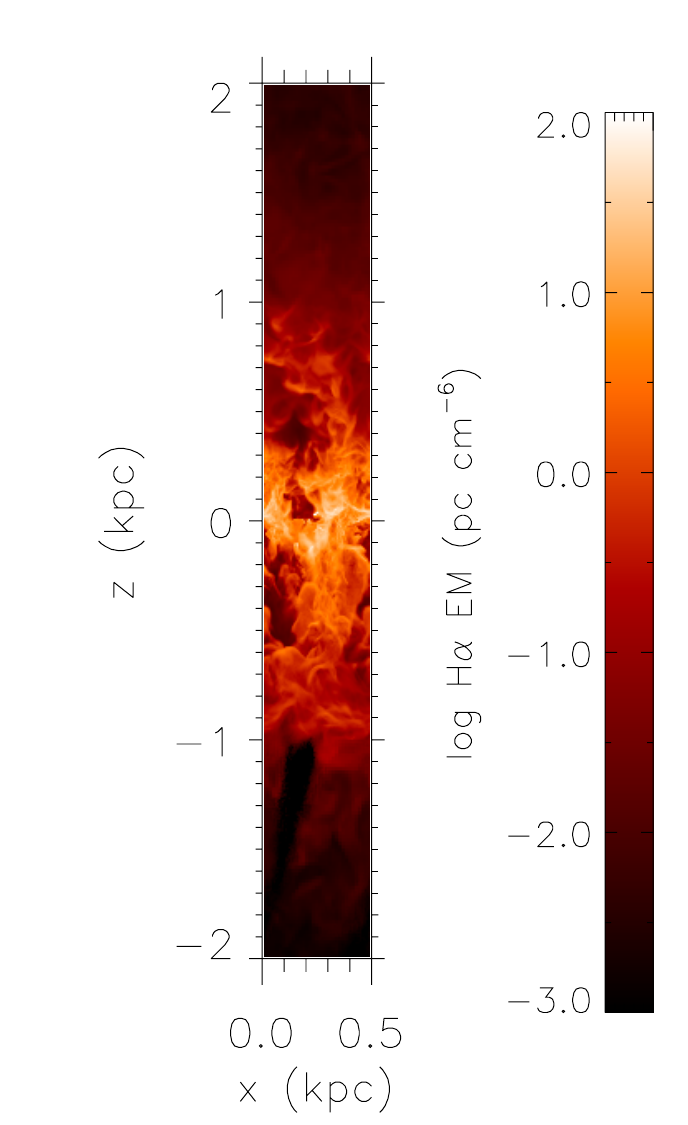}
\includegraphics[scale=0.25]{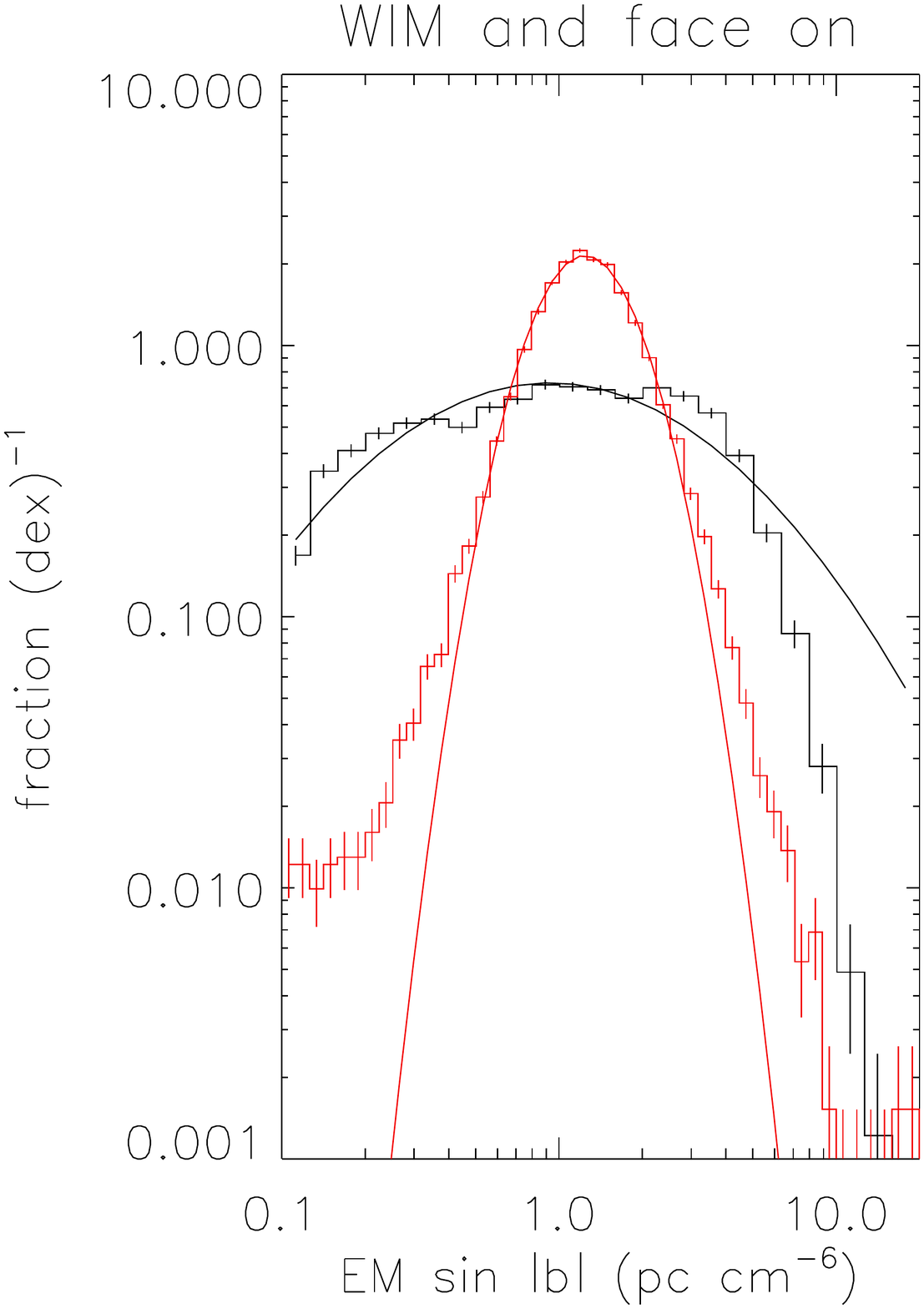}
\caption{\label{fig:ionization} {\em(a)} Edge-on visualizations of
  emission measures ($\int n_e^2 ds$) from a simulation grid described
  by \citet{joung09}. Adapted from Fig.~3 of \citet{wood10}. {\em (b)}
  Emission measure distributions for the Galactic DIG from the WHAM
  survey and from our photoionization simulation. Fraction
  (dex)$^{-1}$ is the fraction of points in each bin divided by the
  logarithmic bin interval. The black histogram is of $\int n_e^2 ds$
  in the simulation, viewed face-on. The red histogram is of EM~$\sin
  |b|$ from the WHAM survey with classical H~{\sc ii} regions and
  sightlines with $|b| < 10^\circ$ removed, from
  \citet{hbk08}. Lognormal fits to each distribution are also
  shown. For the emission measures from the simulations we have
  removed contributions from midplane regions with $|z|<
  150$~pc. 
Adapted from Fig.~6 of \citet{wood10}.  }
\end{figure}

\citet{wood10} compared the distribution of H$\alpha$ emission
measures measured by the Wisconsin H$\alpha$ Mapper
\citep[WHAM][]{haffner03} along lines of sight out of the plane of the
Galaxy to the distribution of simulated H$\alpha$ emission from the
model (see Fig.~\ref{fig:ionization}{\em (a)}).
Figure~\ref{fig:ionization}{\em (b)} shows the
comparison.  They found that the emission measure distribution of the
models was centered around the same value as the observations, but
that it was broader, with more lines of sight having very high or low
emission measures seen in the models.  This suggested that some
additional homogenizing effect might be acting beyond the physics
included in the models.  Magnetic fields could have such an effect, so
we decided to investigate how their inclusion would change our
results.  We here report on our implementation of fields in our models
and some first results, though we only reach preliminary conclusions.

\section{Simulation Techniques}

Our problem requires the computation of flows with Alfv\'en and Mach
numbers greatly exceeding unity in plasmas with a range of plasma
$\beta$ (ratio of thermal to magnetic pressure) extending from well
under to well over unity.  Most linearized Riemann solvers have
substantial stability problems under these conditions.  Therefore we
used the positivity-preserving generalization by \citet{bouchut07} of
an HLL \citep{harten83} approximate Riemann solver with a base
MUSCL-Hancock scheme.  This solver satisfies discrete entropy
inequalities and imposes positivity of density and internal energy.
The three-wave HLL3R version of \citet{bouchut10} is used here, in an
implementation described by \citet{waagan09}.  This solver vastly
improves stability, and thus efficiency, without sacrificing accuracy.

\citet{waagan11} demonstrates that HLL3R performs as well as other HLL
schemes on standard test problems such as shock tubes. 
Figure~\ref{fig:o-t} shows a comparison of solutions of an Orszag-Tang
vortex by two of the standard Flash solvers, Roe and HLLE to the
HLL3R solution.  The accuracy compares to the Roe
solver, while the slightly more diffusive HLLE solver fails to capture
the central structure.  However, the comparison of the performance of
the Roe solver to HLL3R for driven Mach 2 turbulence given in Table~3
of \citet{waagan11} shows that the
Roe solver requires a Courant-Friedrich-Lewy number below 0.2 for
stability, while HLL3R remains stable at 0.8.  Even when the Roe
solver doesn't crash outright, it regularly generates hot zones that
force the Courant timestep to low values.  As a result, the total CPU
time used for this problem is a factor of 9 lower for HLL3R.
Increasing the Mach number to 10 in $\beta = 1$ plasma resulted in a
problem that simply could not be stably computed by the Roe or HLLE
solvers, but was solved by HLL3R. This stability was demonstrated under
even more extreme conditions in the driven MHD turbulence models by 
\citet{brunt10} which reached Mach numbers as high as 20 with $\beta =
1.25 \times 10^{-3}$. 
\articlefigure[scale=0.7]{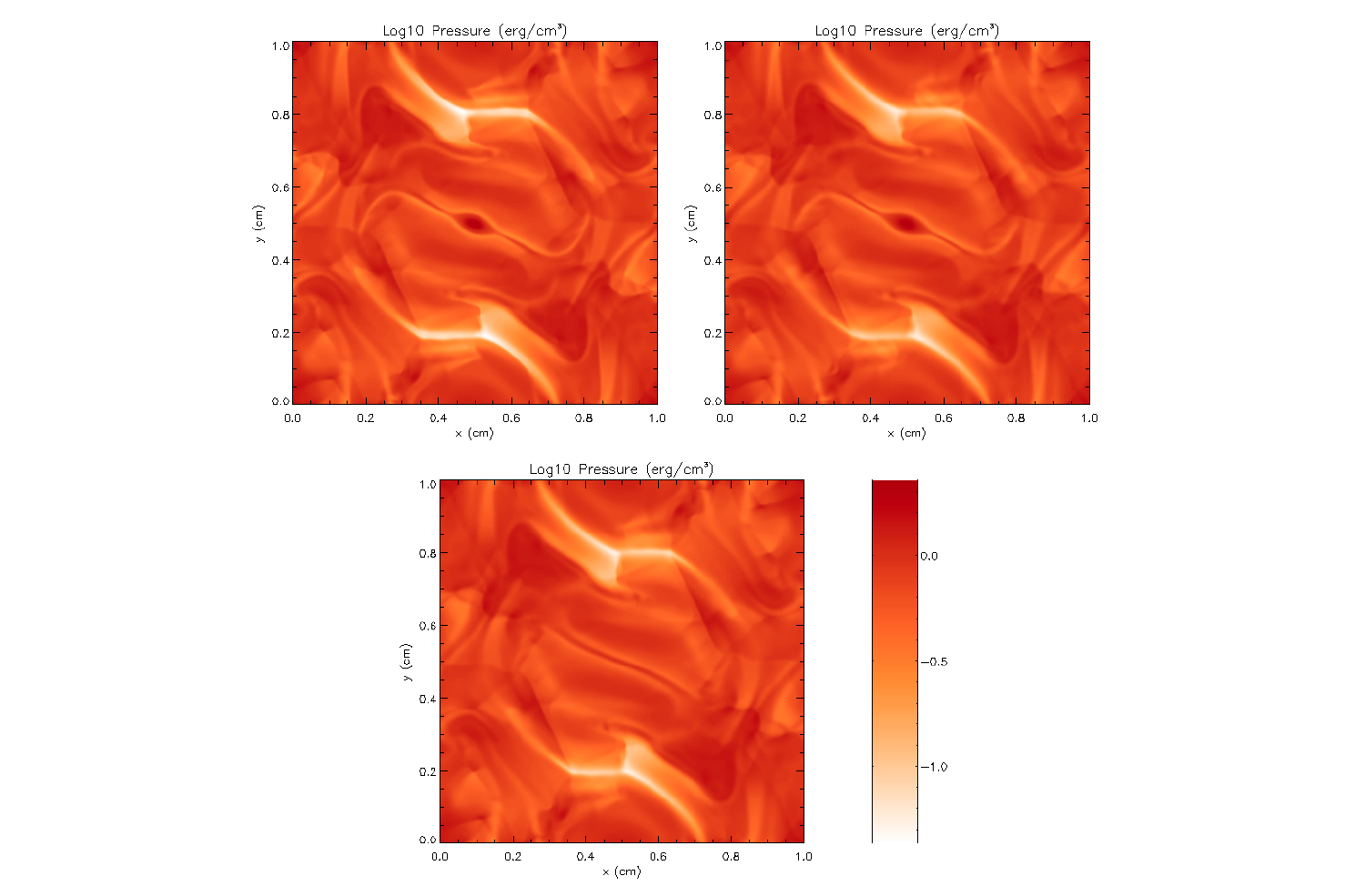}{fig:o-t}{Orszag-Tang
  test. Pressure at time $t = 1$ with resolution $h = 256^{−1}$. Top
  to bottom, left to right: FLASH-Roe, HLL3R, FLASH-HLLE. Figure 4.3
  from \citet{waagan11}}.

Our magnetized, supernova-driven models were initialized with constant
$\beta$ fields having midplane strengths of either 6.5 or 13~$\mu$G.
We found during initial testing that the grid we used in
\citet{joung06} and \citet{joung09} was insufficiently large to
capture the dynamics of the halo, though it sufficed to model the
midplane gas that we focused on in those papers.  Instead, we used $1
\times 1 \times 40$~kpc grids for our standard models.  Also, instead
of initializing an entirely isothermal grid at $10^4$~K, we set the
halo temperature to $10^6$~K at altitudes $|z| > 1$~kpc to avoid very
strong shocks at high altitude during the initial expansion of
supernova blast waves.

\section{Preliminary Results}

\subsection{Vertical Structure}

The magnetic field indeed leads to somewhat smoother flows, with more
filamentary structure seen above a kiloparsec.  However, the biggest
surprise came from examindation of the vertical structure of the halo.
The ever increasing halo temperature seen by \citet{henley10} was
shown to be a numerical artifact.  The constant-pressure vertical
boundary conditions used by \citet{joung06} and \citet{joung09} did
not have a velocity limiter to prevent inflow.  After a hot initial
transient flow from the first supernova explosions in the halo crossed
the boundary, the pressure and temperature were set high.  The gas
then started to fall inward lacking further support, and the boundary
then fed high pressure and temperature continuously on to the grid.
Raising the boundary and setting it to allow only outflow resulted in
much more reasonable behavior, with an appropriately thick disk not
confined by high pressure gas, and a moderate temperature halo, as shown in
Figure~\ref{fig:oscillation}.  In future work we will examine in
detail whether the effective X-ray temperature, H$\alpha$ emission
measure, and H~{\sc i} 21~cm distributions now reproduce the observations.
\articlefigure[scale=0.55]{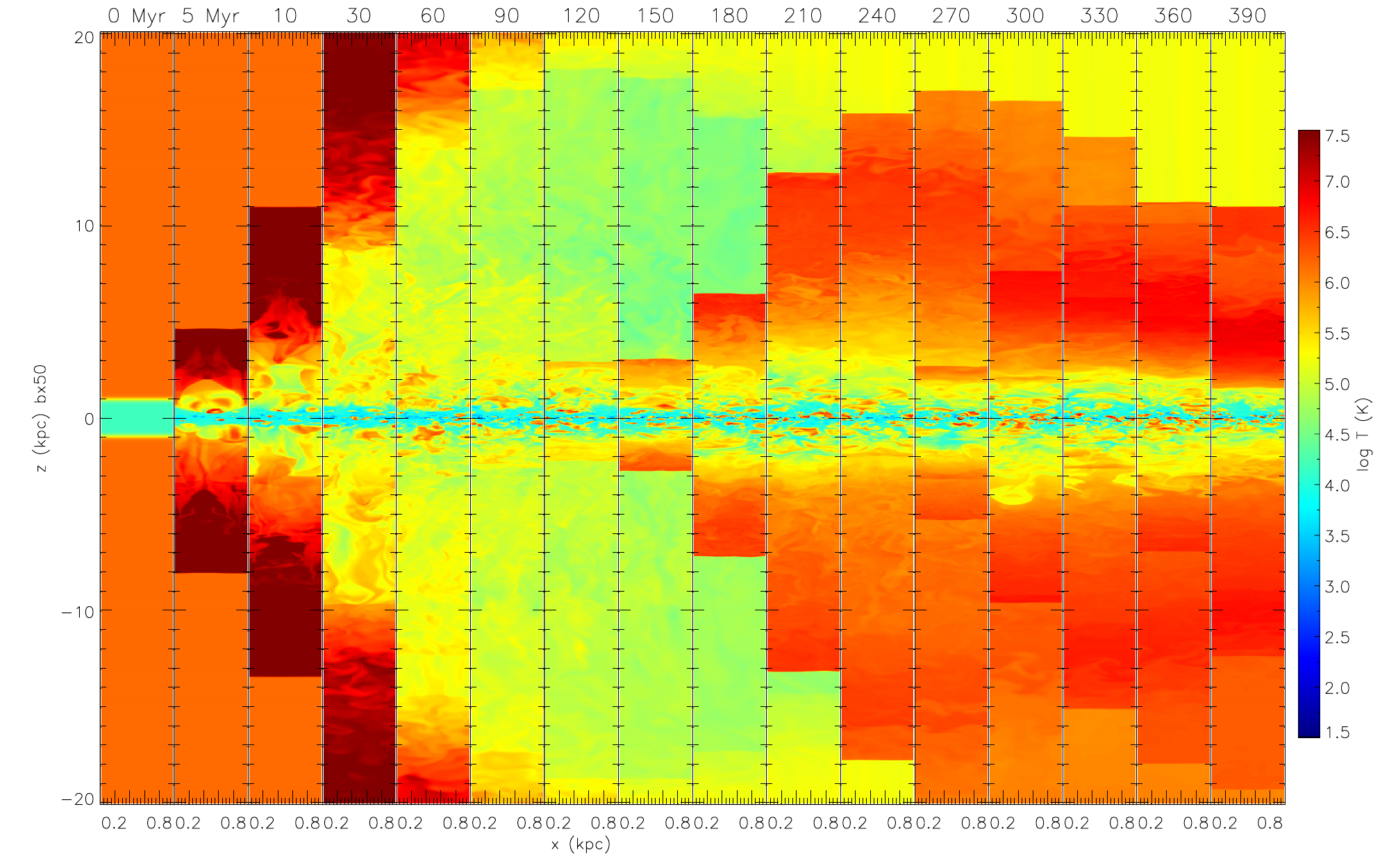}{fig:oscillation}{Slices
  of temperature over time for a model with initial midplane magnetic
  field of 6.5 $\mu$G. The time of each slice in megayears is listed above
  each image. Note that the aspect ratio is not unity; the
  vertical scale of these images is contracted. Adapted from Figure~2
  of \citet{hill11}}

However, we then saw that even in steady state, the halo 
undergoes vertical oscillation on either side of the midplane,
uncorrelated between the upper and lower sides.  Exactly this behavior
was predicted by \citet{walters01} using a one-dimensional model and
proposed as an explanation for the remarkable observation that gas
towards the northern Galactic pole seems to be primarily infalling
while that towards the southern Galactic pole seems to primarily flow
outwards. Although our models reproduce only a small portion of a
galactic disk, we may take from them the qualitative statement that
the upper atmospheres and halos of galaxies are unlikely to be
quiescent, static places, but rather are likely to be in continuous
hypersonic motion, with observable consequences.

\subsection{Dynamos}

The modeling of an idealized turbulent dynamo using HLL3R was already
demonstrated by \citet{waagan11}.  This work was extended to consider
dynamos in accretion flows with application to the formation of the
first stars by \citet{sur10}.  \citet{federrath11} demonstrated that
dynamo action in accretion only occurs for resolutions of at least 30
cells per Jeans length, and does not show convergence out to at least
128 cells per Jeans length, instead showing continuously stronger
growth of rms field with higher resolution and higher effective
Reynolds number \citep{federrath11b}.
\citet{balsara04} had already
shown that supernova-driven turbulence in a periodic box acts as a
turbulent dynamo.  \citet{hill11} shows that this likely holds in a
stratified medium as well, with magnetic field energies initially
varying dependent on their initial values but then appearing to
converge towards values characteristic of field strengths somewhat
under 5~$\mu$G.  However runs of 400~Myr still had not clearly moved
into a steady-state regime, so this conclusion remains preliminary.

\section{Conclusions}

Supernova-driven turbulence appears increasingly likely to be able to
explain observations in H$\alpha$ and X-ray of material above the
plane of the Galaxy, along with many other observations.   Hypersonic
vertical motions of large regions of galactic halos above star-forming
disks appear to be driven by transient variations in energy input in
the much denser disk midplanes of the galaxies, as predicted by
\citet{walters01}. These vertical motions appear to explain at least
some of the discrepancies between models and observations identified
by \citet{henley10} and \citet{wood10}.  Our magnetized models
demonstrated that magnetic fields can be modeled stably even in highly
compressible, highly magnetized flows using the positivity preserving
HLL3R solver described by \citet{waagan09}. They further suggested
that a turbulent dynamo can be driven in a stratified disk, as well as
in a periodic box.

\acknowledgements This work was partly supported by NASA/SAO grant
TM0-11008X and by NSF grant AST-0607512. The software used in this
work was in part developed by the DOE NNSA-ASC OASCR Flash Center at
the University of Chicago.  

\bibliographystyle{asp2010}
\bibliography{maclow}

\end{document}